\documentclass[11pt]{article}
\usepackage{arxiv}
\usepackage{cite}
\usepackage{amsmath,amssymb,amsfonts}
\usepackage{algorithmic}
\usepackage{algorithm}
\usepackage{graphicx}
\usepackage{textcomp}
\usepackage{xcolor}
\usepackage{booktabs}
\usepackage{multirow}
\usepackage{url}
\usepackage{subcaption}
\usepackage{siunitx}

\graphicspath{{figures_v3/}}

\title{MINT: Dynamic-Precision CNN Inference with MSDF Digit-Serial Arithmetic on FPGA}

\usepackage{authblk}

\author[1]{Muhammad Usman}
\author[2]{Malik Zohaib Nisar}
\author[3]{Florian Aschauer}
\author[1]{Dorit Merhof}

\affil[1]{Faculty of Informatics and Data Science, University of Regensburg, 93053 Regensburg, Germany \protect\\ \texttt{\{muhammad.usman,dorit.merhof\}@ur.de}}
\affil[2]{Department of Computer Engineering, Chosun University, Gwangju, Republic of Korea \protect\\ \texttt{zohaib@chosun.ac.kr}}
\affil[3]{Faculty of Electrical and Information Engineering, Ostbayerische Technische Hochschule, Regensburg, Germany  \protect\\ \texttt{florian.aschauer@oth-regensburg.de}}

\begin{document}
\maketitle
\begingroup
\renewcommand\thefootnote{}
\footnotetext{Accepted for publication at IEEE 39th International System-on-Chip Conference (SOCC)}
\endgroup
\begin{abstract}
We present MINT, a dynamic-precision CNN inference accelerator based on left-to-right (LR) arithmetic. LR arithmetic computes in most-significant-digit-first manner and exposes useful partial results early so that the computation can be terminated once the desired precision is achieved. At the core, there is a MSDF serial-parallel inner-product unit, which uses redundant signed-digit representation to compute each convolution window. A budget-constrained greedy search profiles all convolution layers from INT2 to INT7 and selects the lowest precision per layer while constraining total accuracy loss to within 2\% of the INT8 baseline for VGG-16 and ResNet-18 networks. The design is synthesized on a Xilinx Zynq-7020 at \SI{200}{\mega\hertz}, and uses 5.64 average bits for VGG-16 and 6.04 for ResNet-18, while achieving 19.86 GOPS and 29.51 GOPS/W on VGG-16, and 18.86 GOPS and 26.40 GOPS/W on ResNet-18. This corresponds to 32.6\% and 26.0\% higher throughput and 82.10\% and 62.90\% higher energy efficiency than INT8 with only 1.81\% and 1.96\% drops relative to the INT8 baseline. Compared with representative prior FPGA CNN accelerators considered in this study, MINT delivers the highest energy efficiency among the listed VGG-16 and ResNet-18 designs on Zynq-7020 platform.
\end{abstract}


\section{Introduction}
\label{sec:intro}
Convolutional neural networks (CNNs) have achieved remarkable accuracy in image classification, object detection, and segmentation tasks~\cite{zhao2024review}. These networks are often trained in floating point numbers with precision of $32$ bits. However, the inference hardware commonly uses reduced bitwidth fixed point (FXP) arithmetic such as INT8. It is well-established that different DNN layers can operate at different precision without suffering from accuracy loss \cite{rakka2024review}. To this end, several recent works have been presented to accelerate the CNN inference on hardware with flexible data precision to reduce compute time and save energy \cite{wang2019haq}. Most quantized accelerators use conventional Least-Significant-Digit-First (LSDF) arithmetic.  In LSDF datapaths, early termination discards the \emph{most significant} digits, which results in a large error. On the other hand, left-to-right arithmetic (LR)~\cite{ercegovac_lang_book}, operates in the \emph{most-significant-digit-first} (MSDF) order using a redundant signed-digit (SD) representation. Output digits emerge from the most significant position first, so terminating after $P$ digits yields a natural $P$-digit approximation with bounded error $\leq 2^{-P}$. This property enables \emph{zero-overhead dynamic precision}, i.e., the same PE hardware can deliver any precision simply by varying the number of clock cycles, with no additional control logic.

The key contributions of this work are as follows:
\begin{itemize}

\item We propose a dynamic-precision CNN inference architecture based on most-significant-digit-first (MSDF) arithmetic, in which precision is controlled solely by early termination of output digits. Unlike conventional mixed-precision designs, the proposed approach requires no hardware reconfiguration, mode switching, or datapath modification, allowing a single processing element to support precisions from INT2 to INT8 seamlessly.

\item We design a serial-parallel MSDF processing element for convolution inner-product computation using a radix-2 redundant representation and an online adder tree. The resulting FPGA architecture naturally exposes a precision--latency tradeoff through its digit-serial execution.

\item We introduce a budget-constrained greedy precision-assignment algorithm that jointly considers layer sensitivity, MAC distribution, and the MSDF cycle model to determine per-layer precision. Unlike prior mixed-precision methods, the proposed optimization maps precision decisions directly to cycle-level savings enabled by early termination.

\item We evaluate the proposed design on VGG-16 and ResNet-18, demonstrating up to 18.1\%  and 1.82$\times$ improvement in latency and energy efficiency respectively, compared with the INT8 baseline, while keeping the accuracy degradation within a 2\% budget.

\end{itemize}

In the remainder of this paper, the proposed dynamic-precision accelerator is referred to as MINT.
\section{MSDF Arithmetic} \label{sec:background}
MSDF arithmetic process operands digit-by-digit, in most significant
first manner~\cite{ercegovac_lang_book}. To produce the output digit with highest weight first, the algorithm needs to have redundancy in the input for which the signed-digit (SD) are usually employed. In this study, we employ radix-2 SD numbers from the digit set $\{-1, 0, 1\}$, where each digit is encoded as a 2-bit pair $(p,n)$ with value $p - n$. The first output digit is produced after a fixed small delay during which a few input digits are processed. This delay is known as \emph{online delay}, denoted as $\delta$, thereafter, one output digit is produced every clock cycle. For radix-2 serial-parallel MSDF multiplier: $\delta_{\text{mult}} = 2$ and MSDF adder: $\delta_{\text{add}} = 2$.
To compute a full $P$-precision digit serial-parallel multiplication (where $P$ $\in$ $\mathbb{Z}$), the algorithm takes $(2*P) + \delta_{\text{mult}} = 2P + 2$ cycles.

\paragraph{Radix-2 Left-to-Right Serial-Parallel Multiplier (LRM)}
The MSDF multiplier implements radix-2 digit-serial multiplication  using a redundant signed-digit representation. One operand is provided serially while the other is available in parallel. At iteration $j$, an internal residual $w[j]$ is updated using the serial input $x_j$ and parallel input $Y$ as:
\[v[j] = 2w[j] + x_{j+\delta}\cdot Y,\]
followed by digit selection $z_{j+1} = \mathrm{SEL}(v[j])$ and residual update: \[ w[j+1] = v[j] - z_{j+1}.\] After an online delay of $\delta=2$, one output digit is produced per cycle in MSDF order. The redundant representation bounds the residual and eliminates long carry propagation, enabling constant-time digit generation and early termination once sufficient precision is obtained. The block diagram of multiplier is shown in Fig.~\ref{fig:architecture}(c), while detailed algorithms and design considerations can be found in~\cite{usman2023low}.

\paragraph{Radix-2 Left-to-Right Adder (LRA)}
The MSDF adder shown in Fig.~\ref{fig:architecture}(d) consists of two full-adders, local recoding logic, and latches for intermediate carry-related signals. The upper stage processes the current input digits and forms bounded intermediate terms, which are then combined in the lower stage to generate the redundant output pair \(z^{+}\) and \(z^{-}\). The carry information is handled locally through recoding and intermediate storage, thereby avoiding global carry propagation. After the initial online delay, the adder produces one output digit per cycle, enabling efficient integration into reduction-tree accumulation within the proposed architecture. Further details are available in~\cite{dslrcnn}.

\section{Proposed Methodology}
\label{sec:arch}
\subsection{Processing Engine (PE)}
\begin{figure}[t]
    \centering
    \includegraphics[width=0.50\textwidth]{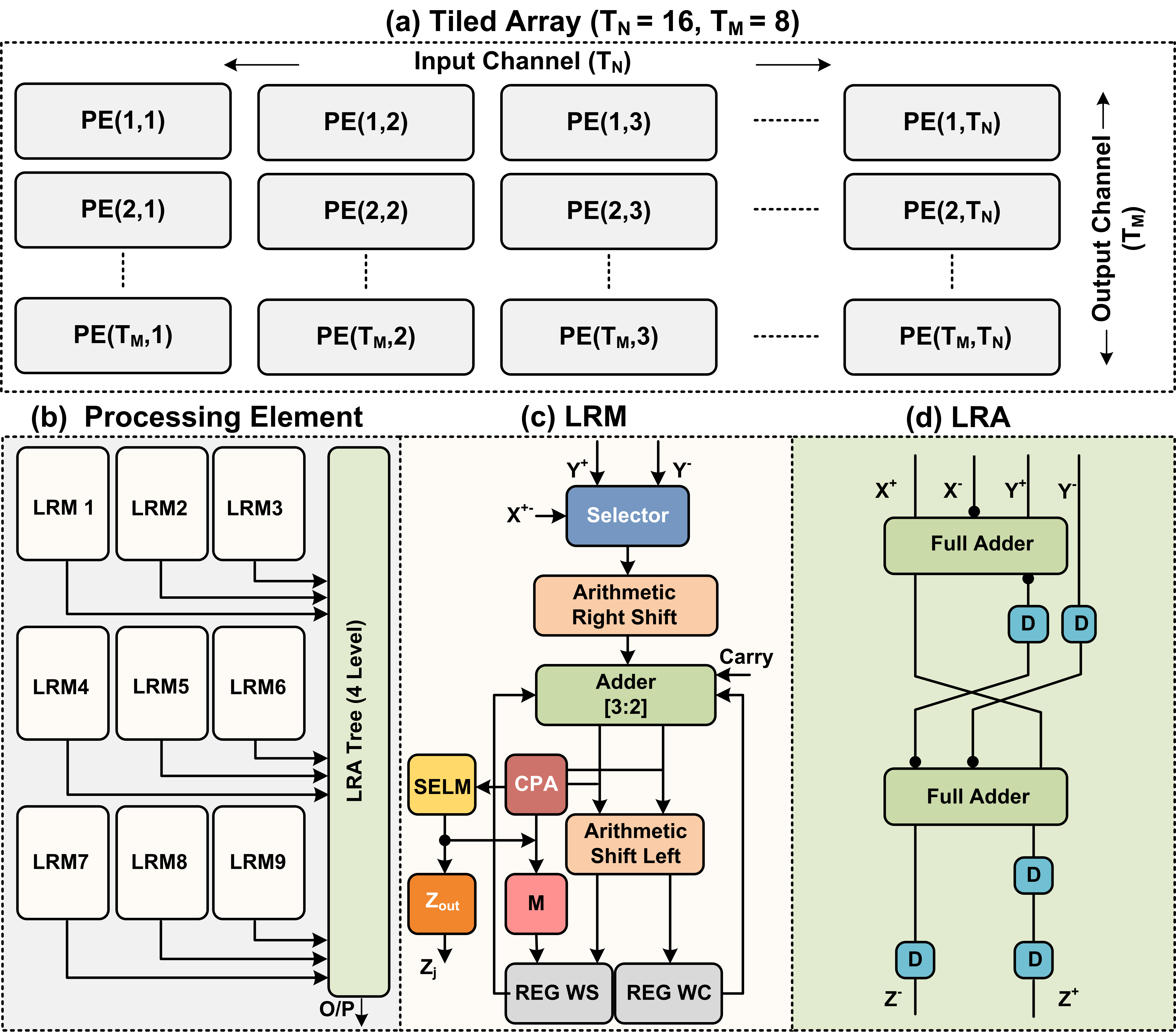}
    \caption{Proposed architecture. (a) Tiled array with input/output channel tiling (b) MSDF PE with serial-parallel multipliers, hierarchical LRA-Tree  (c) LR multiplier (d) LR adder}
    \label{fig:architecture}
\end{figure}

The processing engine (PE) shown in Fig.~\ref{fig:architecture}(b), is essentially a 9-tap inner product $z = \sum_{i=1}^{9} x_i \cdot Y_i$, to compute the $3\times3$ convolution kernel applied at one spatial position. It comprises of 9 serial-parallel MSDF multipliers followed by MSDF adder tree. The serial input $x_i$ is fed MSD-first as SD digit pairs while the parallel input $Y$ stores the filter weight. The 9-tap adder tree sums 9 multiplier outputs through $\lceil\log_2(9)\rceil = 4$ adder stages.
Each stage introduces $\delta_{\text{add}} = 2$ additional cycles.
The total latency for a $P$-digit inner product is:
\begin{equation}
  C(P) = \underbrace{2P + \delta_{\text{mult}}}_{\text{multiplier}} +
         \underbrace{\delta_{\text{add}} \times \lceil\log_2(9)\rceil}_{\text{4-stage adder tree}} = 2P + 10
  \label{eq:cycles}
\end{equation} 

\subsection{PE Array}

As discussed earlier, the digit-serial compute units produce their outputs over multiple clock cycles. Although this increases the computation latency of a single unit, these units occupy a smaller area and exhibit a shorter critical path. As a result, higher throughput can be achieved by instantiating multiple digit-serial units in parallel. The arrangement of several parallel compute units to accelerate CNN processing is commonly referred to as \emph{tiling}. Bit-parallel architectures require higher power, larger area and interconnects, therefore their achievable tiling factor is constrained. Since tiling has a direct impact on overall performance, it is usually selected carefully based on the target model structure and hardware constraints.

In this work, our primary objective is to demonstrate the effectiveness of the proposed MSDF inner-product unit for dynamic-precision computation, rather than to optimize the tiling configuration. Therefore, we adopt a fixed input tiling factor ($T_N$) of 16 and an output tiling factor ($T_M$) of 8, resulting in a total of 128 PEs as depicted in Fig.~\ref{fig:architecture}(a). All PEs operate using a common clock and reset signal. At \SI{200}{\mega\hertz}, the effective throughput at precision $P$ is given by:
\begin{equation}
  \text{Eff.\ GOPS} = 2\times \frac{T_N \times T_M \times LRM_{PE}}{C(P) \times 5\,\text{ns}}
  \label{eq:gops}
\end{equation}

\subsection{Methodology Overview}
\begin{figure}[t]
  \centering
  \includegraphics[width=0.4\textwidth]{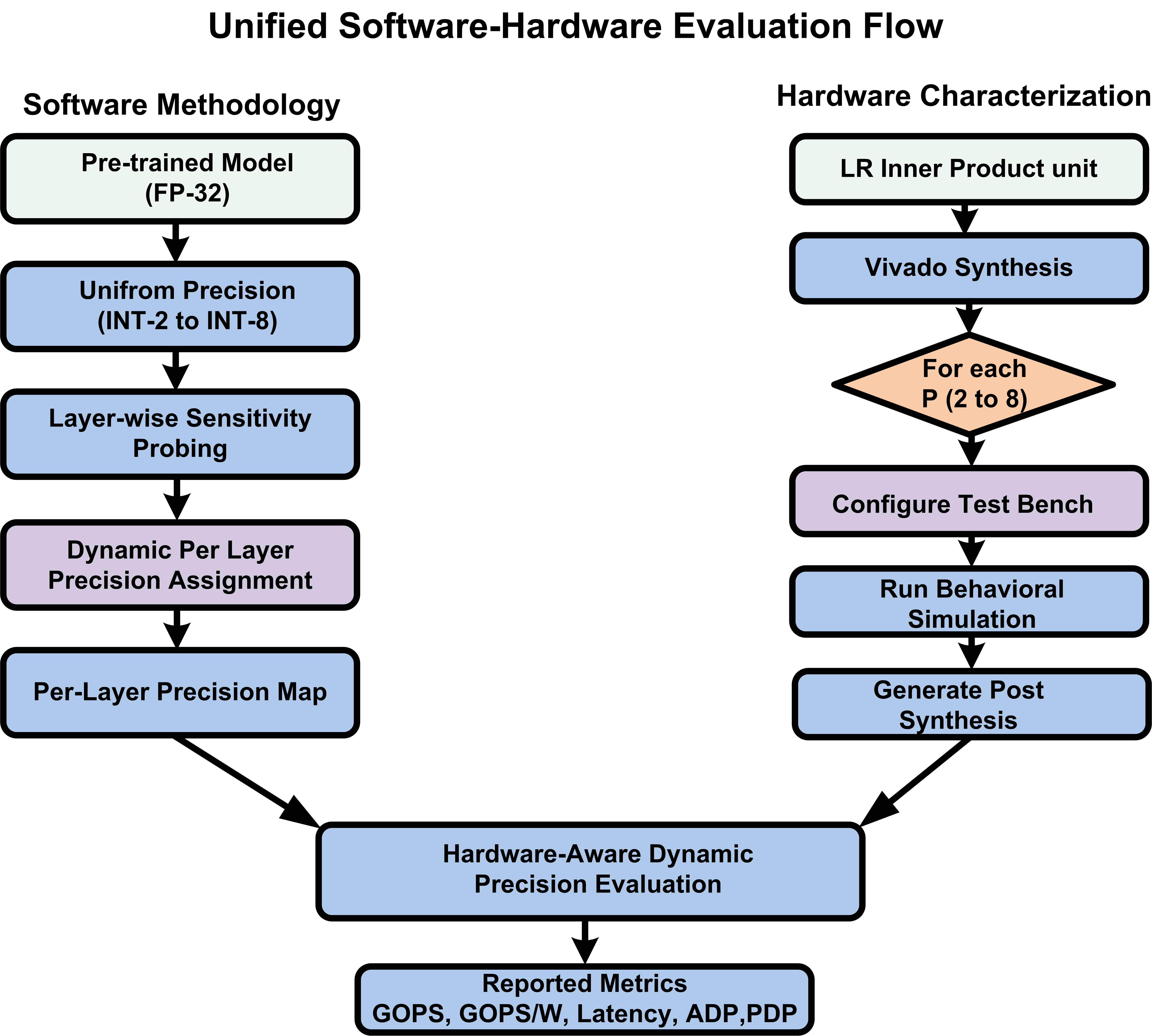}
    \caption{Unified software-hardware evaluation flow.}
  \label{fig:flowchart}
\end{figure}

The complete evaluation flow of the proposed framework is summarized in Fig.~\ref{fig:flowchart}. The software branch starts from a pre-trained FP32 network and performs uniform-precision ablation from INT2 to INT8, followed by layer-wise sensitivity probing. Based on these results, a budget-constrained greedy algorithm assigns a dynamic precision to each convolution layer. In parallel, the hardware branch characterizes the proposed MSDF inner-product processing element using Vivado synthesis, post-synthesis simulation, and Switching Activity Interchange Format (SAIF) based switching-activity analysis across the supported precisions. Finally, the model-level precision map and the hardware-level characterization are combined to obtain hardware-aware results, including throughput, energy efficiency, and latency.

\subsection{Budget-Constrained Greedy Precision Assignment}

A key challenge in mixed-precision quantization is that the accuracy loss introduced by different layers does not accumulate independently. It is therefore important to study the compounding effect on accuracy after selecting a certain precision of each layer. To this end, we use a budget-constrained greedy precision assignment strategy, summarized in Algorithm~\ref{alg:probing}. The method starts from a uniform baseline precision $B$ and incrementally reduces the precision while enforcing the user-defined threshold on accuracy $\Delta_{\max}$.

The sensitivity of each convolution layer is first measured by quantizing one layer at a time to a candidate precision $p \in \mathcal{P}$ while keeping all other layers at $B$. This yields the single-layer accuracy drop $d_{\ell,p}$. The candidate pairs $(\ell,p)$ are then ranked using
\begin{equation}
    \eta'_{\ell,p} = \frac{\text{cycle\_saving}_{\ell,p} \times \text{mac\_frac}_{\ell}}{d_{\ell,p} + \epsilon},
\end{equation}
where $\text{cycle\_saving}_{\ell,p}$ denotes the reduction in execution cycles, $\text{mac\_frac}_{\ell}$ is the MAC fraction of layer $\ell$, and $\epsilon$ is a small constant. The candidates are applied greedily in descending order of $\eta'_{\ell,p}$, with full-network evaluation after each update to account for inter-layer interactions. The efficiency score  uses the precision-dependent cycle count and layer MAC fraction from the hardware cost model.

\begin{algorithm}[t]
\caption{Budget-Constrained Greedy Precision Assignment}
\label{alg:probing}
\begin{algorithmic}[1]
\REQUIRE Pre-trained model $\mathcal{M}$, baseline precision $B{=}8$,
         probe set $\mathcal{P}{=}\{2,\ldots,7\}$, budget $\Delta_{\max}$
\ENSURE Per-layer precision map $\pi$
\STATE $\mathrm{acc}_B \leftarrow \mathrm{Eval}(\mathcal{M}, B)$
\STATE $\pi[\ell] \leftarrow B$ for all conv layers $\ell$
\STATE \textit{// Phase 1: Single-layer sensitivity probing}
\FOR{each conv layer $\ell$}
  \FOR{each $p \in \mathcal{P}$}
    \STATE Quantize only $\ell$ to $p$ bits, others at $B$
    \STATE $d_{\ell,p} \leftarrow \mathrm{acc}_B - \mathrm{Eval}(\mathcal{M}_{\ell\rightarrow p})$
  \ENDFOR
\ENDFOR
\STATE \textit{// Phase 2: Greedy assignment}
\STATE Sort candidates $(\ell,p)$ by efficiency
\STATE \hspace{1em} $\eta_{\ell,p} \leftarrow
\frac{\mathrm{cycle\_saving}_{\ell,p}\times \mathrm{mac\_frac}_{\ell}}
{d_{\ell,p}+\epsilon}$
\FOR{each candidate $(\ell,p)$ in descending $\eta$}
  \STATE $\pi_{\mathrm{old}} \leftarrow \pi[\ell]$
  \STATE $\pi[\ell] \leftarrow p$ \textit{(tentative)}
  \STATE $\mathrm{acc} \leftarrow \mathrm{Eval}(\mathcal{M}_{\pi})$
  \IF{$\mathrm{acc}_B - \mathrm{acc} \leq \Delta_{\max}$}
    \STATE Accept: keep $\pi[\ell] = p$
  \ELSE
    \STATE Reject: restore $\pi[\ell] \leftarrow \pi_{\mathrm{old}}$
  \ENDIF
\ENDFOR
\end{algorithmic}
\end{algorithm}

\section{Experimental Setup}
\label{sec:setup}

\subsection{Software Environment}
We evaluate VGG-16 and ResNet-18 using pre-trained ImageNet \cite{imagenet} weights. Post-training quantization is performed in PyTorch with per-channel symmetric quantization and histogram-based calibration. Weights and activations are quantized to $P$ bits using a uniform quantizer with scale factor $s = x_{\max}/(2^{P-1}-1)$, emulating MSDF output-digit termination by retaining the first $P$ most-significant digits. Accuracy is measured on the Imagenette-320 validation set (3,925 images, 10 classes). Only Conv2d and linear layers are quantized in simulation whereas the remaining layers remain in FP32. As a result, the reported accuracy represents an upper bound on hardware performance.

\subsection{Hardware Platform}
Synthesis is done on Xilinx Zynq-7020 (xc7z020clg484-1, \SI{28}{\nano\metre}) device in Vivado 2024.1 with a \SI{5}{\nano\second} clock constraint (\SI{200}{\mega\hertz}). All multiply-accumulate operations are implemented using LUTs and flip-flops. Power is characterized in Vivado using SAIF-based switching activity annotation. For each precision level, the testbench is simulated, switching activity is back-annotated to the implemented design, and dynamic and static power are extracted. In evaluation, batch normalization is folded into the weights, ReLU/pooling/skip latency is neglected, and fully connected layers are excluded from the PE-array model. PyTorch accumulation remains in FP32, so the reported accuracy is an upper bound. Throughput and power evaluations assume a PE-array utilization factor $\eta = 0.88$, activity factor of $\alpha = 0.50$, and a frequency fixed by the INT8 critical path. The device static power (\SI{105}{\milli\watt}) is counted once at the device level.

\section{Results and Discussion}
\label{sec:results}

\subsection{Uniform Precision Accuracy}

\begin{figure}[t]
  \centering
  \includegraphics[width=0.95\columnwidth]{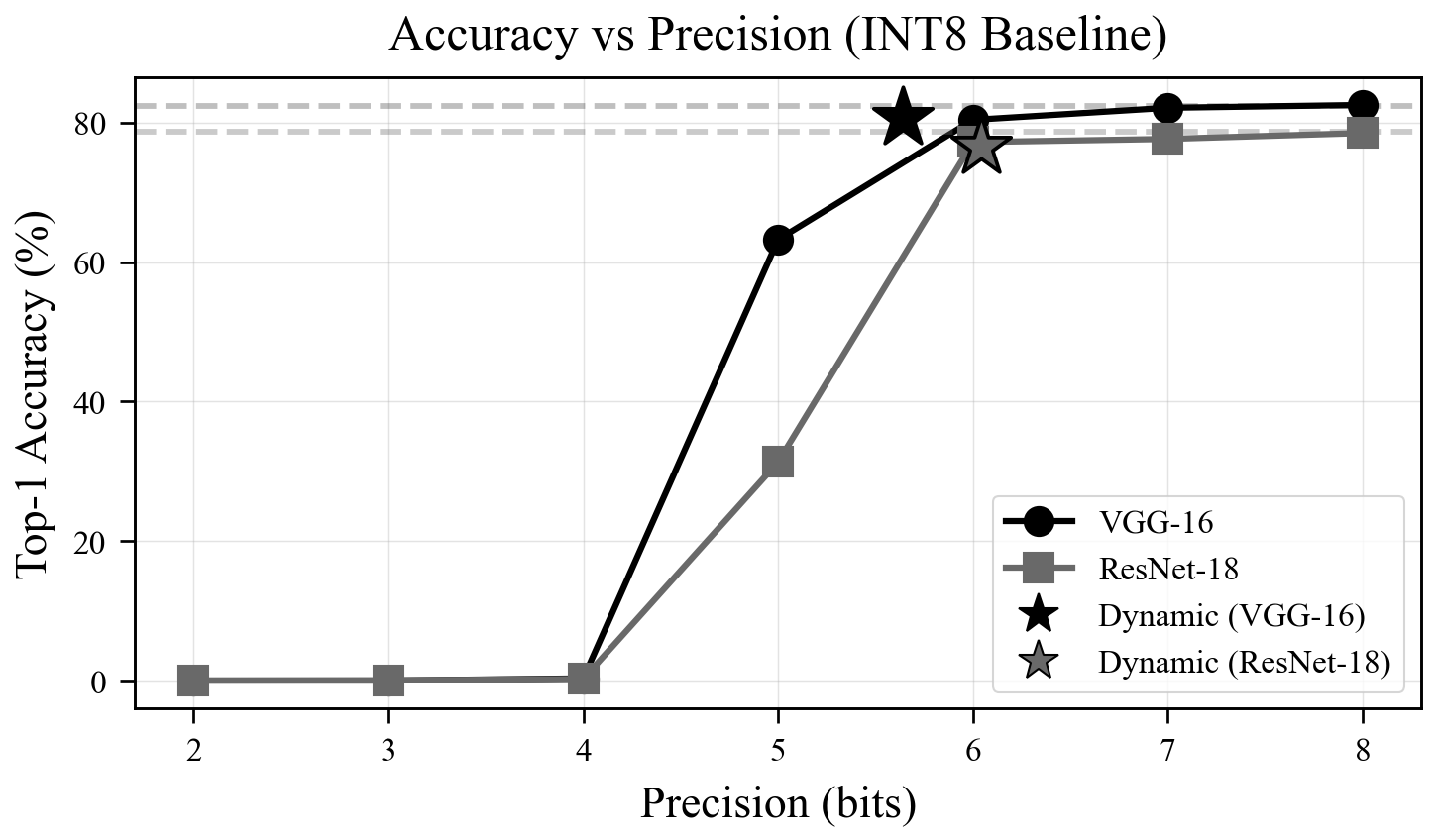}
  \caption{Top-1 accuracy vs.\ uniform precision on Imagenette-320
  for VGG-16 and ResNet-18.  A sharp accuracy cliff is visible between
  INT4 and INT5 for both networks.}
  \label{fig:accuracy}
\end{figure}

Fig.~\ref{fig:accuracy} shows the Top-1 accuracy of VGG-16 and ResNet-18 under uniform quantization from INT2 to INT8. Both networks exhibit a clear accuracy cliff between INT4 and INT5, indicating that INT4 is too aggressive for uniform deployment on Imagenette. VGG-16 improves from \SI{0.13}{\percent} at INT4 to \SI{63.85}{\percent} at INT5, while ResNet-18 improves from \SI{0.28}{\percent} to \SI{34.42}{\percent}. At INT6 and above, both networks recover most of their FP32 accuracy. For VGG-16, INT8 slightly exceeds the FP32 baseline, likely due to quantization regularization effect.

Table~\ref{tab:cycles} summarizes the impact of precision scaling on throughput, hardware savings, and classification accuracy.
\begin{table}[!ht]
\centering
\caption{Uniform quantization accuracy and throughput trade-off (128 PEs, \SI{200}{\mega\hertz})}
\label{tab:cycles}
\setlength{\tabcolsep}{3pt}
\scriptsize
\resizebox{\columnwidth}{!}{%
\begin{tabular}{lccccccc}
\toprule
& & & & \multicolumn{2}{c}{VGG-16} & \multicolumn{2}{c}{ResNet-18} \\
\cmidrule(lr){5-6} \cmidrule(lr){7-8}
Prec. & $C(P)$ & Eff.~GOPS & Savings & Top-1 & $\Delta$ & Top-1 & $\Delta$ \\
\midrule
FP32 & -- & --   & --    & 82.39 & --    & 78.75 & --    \\
INT8 & 26 & 15.0 & 0\%   & 82.55 & $\sim$0  & 78.52 & 0.27  \\
INT7 & 24 & 16.8 & 7.7\% & 82.11 & 0.28  & 77.76 & 0.99  \\
INT6 & 22 & 19.0 & 15.4\%& 80.71 & 1.68  & 76.13 & 2.62  \\
INT5 & 20 & 21.7 & 23.1\%& 63.85 & 18.55 & 34.42 & 44.33 \\
INT4 & 18 & 25.0 & 30.8\%& 0.13  & 82.27 & 0.28  & 78.47 \\
INT3 & 16 & 29.3 & 38.5\%& 0.00  & 82.39 & 0.00  & 78.75 \\
INT2 & 14 & 34.8 & 46.2\%& 0.00  & 82.39 & 0.00  & 78.75 \\
\bottomrule
\end{tabular}%
}
\end{table}

\subsection{Per-Layer Sensitivity and Dynamic Precision Assignment}

\begin{figure}[t]
  \centering
    \includegraphics[width=0.5\textwidth]{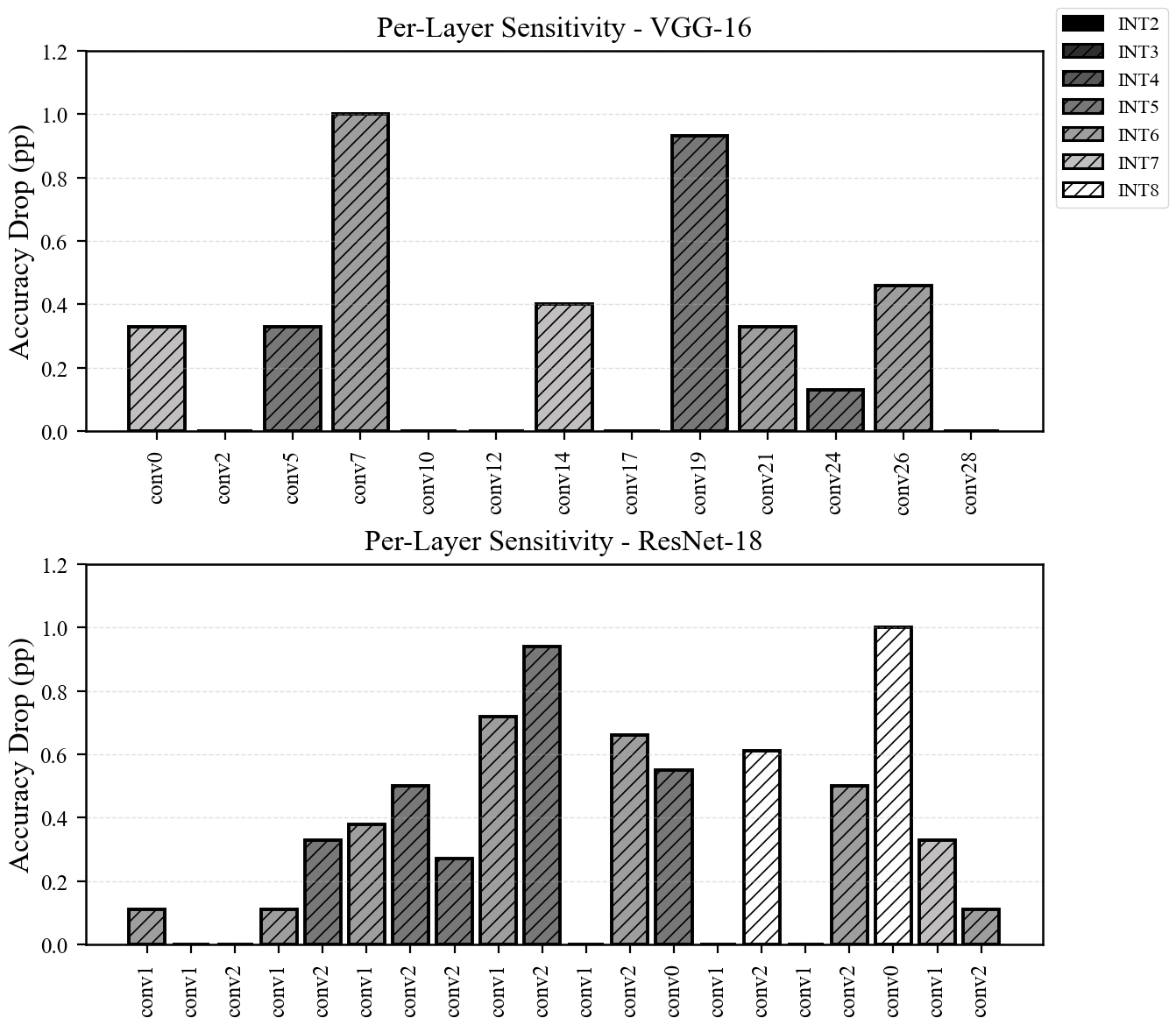}
  \caption{Per-layer sensitivity heatmaps.  Darker cells indicate larger
  accuracy drops.  Layers with low sensitivity at reduced precisions are
  candidates for early termination.}
  \label{fig:sensitivity}
\end{figure}

To understand the limitation of uniform quantization, Fig.~\ref{fig:sensitivity} shows the accuracy drop obtained by quantizing one layer at a time while keeping all other layers at INT8. The sensitivity is highly non-uniform across layers, with only a subset of layers showing severe degradation at low precision. This motivates a layer-wise precision assignment instead of a uniform precision across the full network.

Using this sensitivity profile, the budget-constrained greedy search is applied with $\Delta_{\max}=2\%$. Table~\ref{tab:dynamic} summarizes the resulting mixed-precision configurations. For VGG-16 and ResNet-18, the average precision is reduced to 5.64 and 6.04 bits, respectively, yielding cycle savings of \SI{18.1}{\percent} and \SI{15.1}{\percent}. This translates to throughput improvements of 32.6\% and 26.0\%, GOPS/W improvements of 82.0\% and 62.9\%, and PDP reductions of 45.1\% and 38.6\%, while keeping the accuracy drop within the \SI{2}{\percent} budget.

\begin{table}[t]
\centering
\caption{Dynamic Precision Assignment Results ($\Delta_{\max} = 2\%$)}
\label{tab:dynamic}
\setlength{\tabcolsep}{2pt}
\footnotesize
\resizebox{\columnwidth}{!}{%
\begin{tabular}{llccccccccc}
\toprule
Network & Method & Bits & Top-1 & Drop vs.  & GOPS & GOPS/W & PDP & Cycle & GOPS/W & PDP \\
        &        &      &       &  INT8(\%) &      &        & (nJ) & Savings & Impr. & Red. \\
\midrule
\multirow{3}{*}{VGG-16}
  & FP32    & 32   & 82.39 & ---    & ---  & ---  & ---   & ---   & ---   & ---   \\
  & INT8    & 8    & 82.55 & $-$0.15 & 15.0 & 16.2 & 125.1 & 0\%   & 0\%   & 0\%   \\
  & Dynamic & 5.64 & 80.74 & 1.81   & 19.9 & 29.5 & 68.7  & 18.1\% & 82.0\% & 45.1\% \\
\midrule
\multirow{3}{*}{ResNet-18}
  & FP32    & 32   & 78.75 & ---    & ---  & ---  & ---   & ---   & ---   & ---   \\
  & INT8    & 8    & 78.52 & 0.23   & 15.0 & 16.2 & 125.1 & 0\%   & 0\%   & 0\%   \\
  & Dynamic & 6.04 & 76.56 & 1.96   & 18.9 & 26.4 & 76.8  & 15.1\% & 62.9\% & 38.6\% \\
\bottomrule
\end{tabular}%
}
\end{table}

\begin{figure}[t]
  \centering
  \includegraphics[width=0.85\columnwidth]{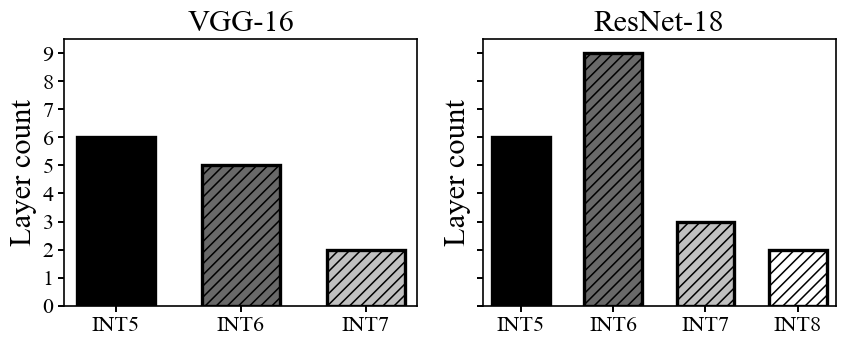}
  \caption{Distribution of assigned precisions across convolutional
  layers for VGG-16 and ResNet-18 under the budget-constrained
  dynamic assignment.}
  \label{fig:precision_dist}
\end{figure}

Fig.~\ref{fig:precision_dist} shows the distribution of assigned
precisions.  VGG-16 assigns most of its compute-heavy mid-network
layers to INT5, while ResNet-18 uses a broader mix of INT5--INT7.

\subsection{Hardware Implementation}
\label{sec:hardware}

Table~\ref{tab:synth} reports the per-PE synthesis results across all
supported precisions.  Because the MSDF PE processes all precisions
with the same hardware, these figures represent the \emph{fixed} INT8
PE resource cost.  Reduced-precision operation saves only cycles
and power, not area.

\begin{table}[t]
\centering
\caption{Per-PE Synthesis Results (Zynq-7020, \SI{28}{\nano\metre})}
\label{tab:synth}
\setlength{\tabcolsep}{4pt}
\begin{tabular}{lccccc}
\toprule
Prec. & LUTs & FFs & Slices & $f_{\max}$ (MHz) & $P_{\mathrm{dyn}}$ (mW) \\
\midrule
INT2 & 190 & 141 &  68 & 240 & 2.7  \\
INT3 & 274 & 202 &  91 & 231 & 4.3  \\
INT4 & 357 & 263 & 113 & 222 & 5.9  \\
INT5 & 449 & 331 & 142 & 214 & 7.5  \\
INT6 & 540 & 399 & 170 & 206 & 9.1  \\
INT7 & 636 & 474 & 200 & 202 & 10.7 \\
INT8 & 732 & 548 & 230 & 200 & 12.3 \\
\bottomrule
\end{tabular}
\end{table}

The INT8 PE occupies 732~LUTs and 548~FFs, meeting timing at \SI{200}{\mega\hertz}. Extrapolating this fixed per-PE cost to the modeled 128-PE array gives an estimated full-design logic utilization of about 94~K LUTs (93{,}696 LUTs), 70.1~K FFs, and 29.4~K slices. Because dynamic precision reuses the same PE array without hardware reconfiguration, this logic cost remains unchanged across precision settings.

\subsection{Throughput and Latency}

\begin{figure}[t]
  \centering
  \includegraphics[width=0.95\columnwidth]{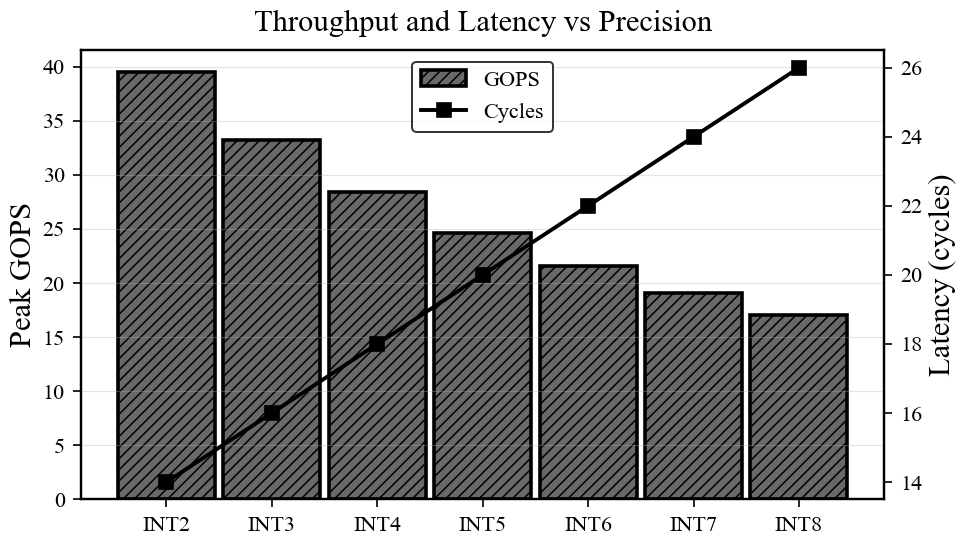}
  \caption{Effective throughput (GOPS) and per-MAC latency vs.\
  precision.  Dynamic operation (marked) achieves throughput between
  INT5 and INT6 levels.}
  \label{fig:throughput}
\end{figure}

Fig.~\ref{fig:throughput} shows throughput and per-MAC latency as a
function of precision for the 128-PE array.
At dynamic precision, the number of average cycle drops
to 21.3 and 22.1 for VGG-16 and ResNet-18 respectively from the INT8 baseline of 26, translating to effective throughput values of \SI{19.9}{GOPS} and \SI{18.9}{GOPS} respectively.

\subsection{Energy Efficiency and Power Analysis}
\begin{figure}[t]
  \centering
  \includegraphics[width=0.95\columnwidth]{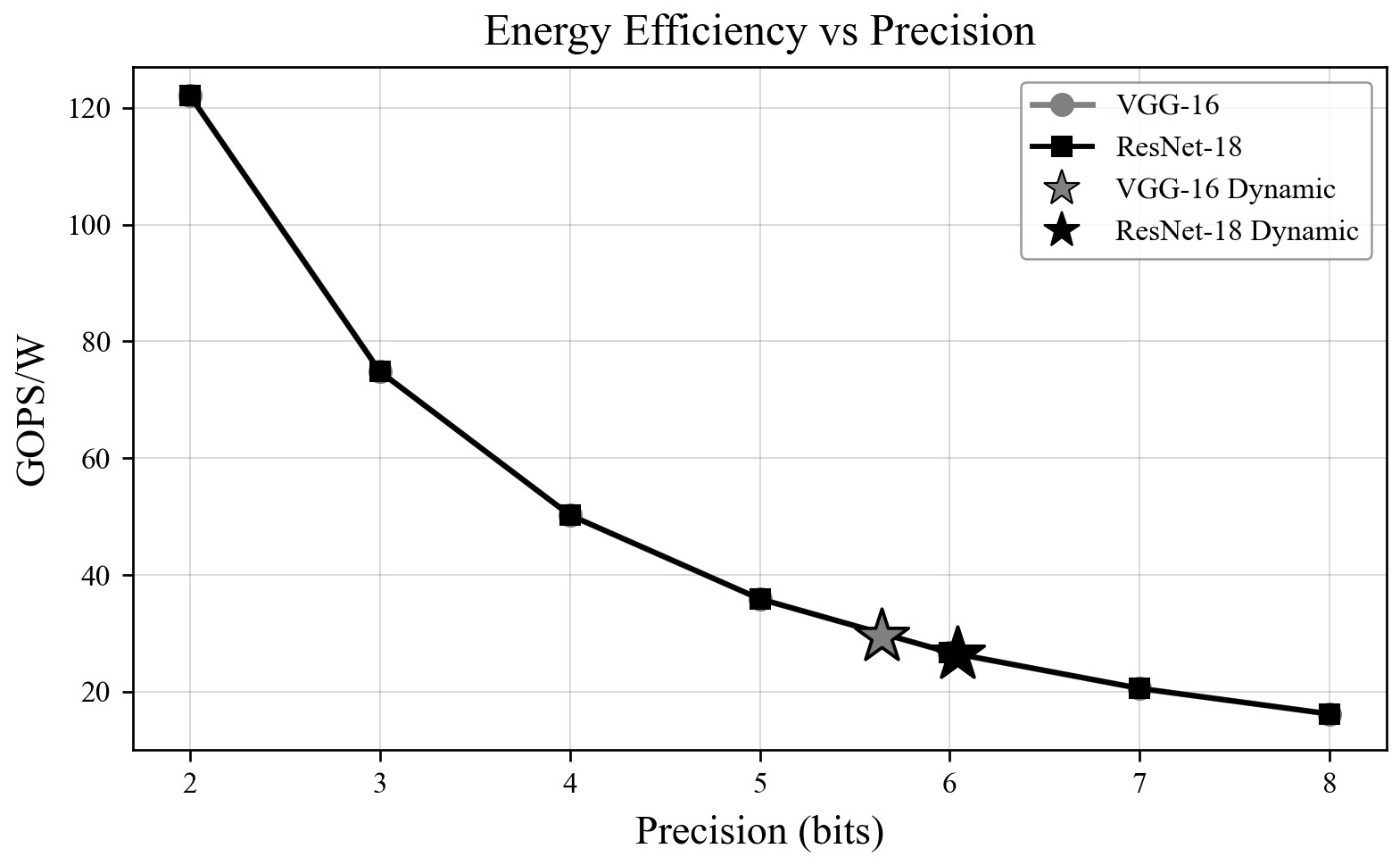}
  \caption{Energy efficiency (GOPS/W) and array power vs.\ precision.
  Dynamic operation achieves \SI{29.5}{GOPS/W} (VGG-16) and
  \SI{26.4}{GOPS/W} (ResNet-18).}
  \label{fig:energy}
\end{figure}

Dynamic operation achieves \SI{29.5}{GOPS/W} for VGG-16, a
$1.82\times$ improvement over the INT8 baseline (\SI{16.2}{GOPS/W}).
The savings arise from two compounding effects: (i)~fewer cycles
per PE reduce execution time, and (ii)~lower-precision operation
reduces dynamic power per cycle. The energy efficiency across different precisions is shown in Fig.~\ref{fig:energy}.

\subsection{Comparison with Previous Works}
\renewcommand{\arraystretch}{1.1}
\begin{table*}[!ht]
\centering
\caption{Comparison with prior FPGA CNN accelerators (best results indicated in bold).}
\label{tab:prevwork}
\setlength{\tabcolsep}{14pt}
\scriptsize
\begin{tabular}{llccccc}
\toprule
Model & Design & Device & Freq. (MHz)$\uparrow$ & Logic Util.$\downarrow$ & GOPS$\uparrow$ & GOPS/W$\uparrow$\\
\midrule
\multirow{6}{*}{VGG-16}
  & OPU~\cite{yu2019opu}     & Zynq XC7Z100 & 200 & 154.5 K      & 397    & 21.45 \\
 & Caffeine~\cite{zhang2016caffeine}& VX690T       & 150 & -- & \text{636}    & 24.46   \\
 & TNNLS'21~\cite{huang2021fpga} & VX980T       & 150 & 335 K & \textbf{1000}    & 14.36   \\
 & TCAD'24~\cite{10312773}& XCVU9P       & \textbf{430} & \textbf{93 K} & \text{711}    & 18.91   \\
  & MINT (INT8)       & Zynq-7020    & 200 & 94 K    & 15.0   & 16.2  \\
 & \text{MINT (Dynamic)} & \text{Zynq-7020} & \text{200} & \text{94 K} & \text{19.9} & \textbf{29.5} \\
\midrule
\multirow{6}{*}{ResNet-18}
 & NEURAghe~\cite{meloni2018neuraghe}& Zynq Z7045    & 140 & 100 K   & 58    & 5.8   \\
 & TCAS'21~\cite{xie2021efficient}  & Arria10 SX660 & 170 & 102.6 K & 89.29 & 19.41 \\
 & ECHO~\cite{ibrahim2024echo}      & VU3P       & 100 & 315 K   & \text{123}   & 21.58 \\
 & TCAD'20~\cite{liang2019evaluating} & XC7Z045 & 200 & 100.2 K  & \textbf{124.90}& 17.09 \\
 & MINT (INT8)       & Zynq-7020     & 200 & 94 K    & 15.0  & 16.2  \\
 & \text{MINT (Dynamic)} & \text{Zynq-7020} & \text{200} & \textbf{94 K} & \text{18.9} & \textbf{26.4} \\
\bottomrule
\end{tabular}
\end{table*}
A comparison of MINT with representative prior FPGA CNN accelerators is presented in Table~\ref{tab:prevwork}. The listed designs target VGG-16 and ResNet-18 workloads, although they differ in FPGA platform, arithmetic style, and optimization objective. The accelerator performance is calculated using the relation $\mathrm{Performance}=\mathrm{Ops}/t$, where $\mathrm{Ops}$ is the total number of operations performed and $t$ is the execution time required to process a given workload. The energy efficiency is then obtained as $\mathrm{GOPS/W}=\mathrm{Performance}/\mathrm{Power}$. 

For the VGG-16 workload, it can be observed from Table~\ref{tab:prevwork} that prior throughput-oriented designs such as OPU~\cite{yu2019opu}, Caffeine~\cite{zhang2016caffeine}, TNNLS’21~\cite{huang2021fpga}, and TCAD’24~\cite{10312773}, achieve higher raw performance in terms of GOPS than MINT. However, MINT in dynamic mode achieves the highest energy efficiency among the listed VGG-16 implementations, reaching \SI{29.5}{GOPS/W}. This corresponds to improvements of 37.53\%, 20.61\%, 105.43\%, and 56.00\% compared with OPU, Caffeine, TNNLS’21, and TCAD’24, respectively. In addition, relative to the fixed INT8 MINT baseline, the dynamic mode improves throughput from \SI{15.0}{GOPS} to \SI{19.9}{GOPS} and improves energy efficiency from \SI{16.2}{GOPS/W} to \SI{29.5}{GOPS/W}, corresponding to gains of 32.67\% and 82.10\%, respectively.

For the ResNet-18 workload, MINT in dynamic mode also achieves the best energy efficiency among the listed designs, reaching 26.4~GOPS/W. Compared with the reported NEURAghe~\cite{meloni2018neuraghe}, TCAS’21~\cite{xie2021efficient}, ECHO~\cite{ibrahim2024echo}, and TCAD’20~\cite{liang2019evaluating}, implementations, this corresponds to 4.55$\times$, 1.36$\times$, 1.22$\times$, and 1.54$\times$ higher energy efficiency, respectively. Although the raw throughput of MINT remains lower than that of the highest-throughput reference implementations, the results show that the proposed dynamic-precision MSDF architecture provides a strong trade-off between resource usage and energy efficiency on a small Zynq-7020 platform.

\section{Limitations and Future Work}


The current implementation validates the proposed dynamic-precision digit-serial
computing approach, but throughput remains limited by the multi-cycle MSDF
multiplier and online adder tree. As a result, the architecture demonstrates the
feasibility and efficiency benefits of dynamic precision, but it is not yet a
fully throughput-optimized real-time accelerator.

Future work will focus on reducing the initiation interval through deeper
digit-level pipelining, larger PE-array scaling, improved scheduling, and more
aggressive early termination of low-significance digits.

\section{Conclusion}
\label{sec:conclusion}

Using a budget-constrained greedy assignment algorithm, the accelerator achieves \SI{18.1}{\percent} cycle savings for VGG-16 (5.64 average bits) and \SI{15.1}{\percent} for ResNet-18 (6.04 average bits) under a \SI{2}{\percent} accuracy-drop budget. Compared with the fixed INT8 baseline, the proposed approach improves energy efficiency by up to \SI{82.0}{\percent} and reduces PDP by up to \SI{45.1}{\percent}, while maintaining the accuracy loss within the target budget, indicating that MINT is a strong candidate for energy-efficient precision-adaptive CNN inference.

\section*{Acknowledgement}\label{sec:Ack}
This work was supported by the German Research Foundation
(Deutsche Forschungsgemeinschaft, DFG) under project number `573796083'.


\bibliographystyle{IEEEtran}
\bibliography{references}

\end{document}